\documentclass[12pt]{article}
\begin{document}
\title{ON THE RELATION BETWEEN THE SLOPES OF DIFFRACTION CONE IN
SINGLE DIFFRACTION DISSOCIATION AND ELASTIC
SCATTERING}
\author{ {A.A.~Arkhipov}\\
\it{Institute for High Energy Physics}\\
\it{Protvino, 142284 Moscow Region, Russia}}
\date{}
\maketitle
\def\be{\begin{equation}}
\def\ee{\end{equation}}
\def\ber{\begin{eqnarray}}
\def\eer{\end{eqnarray}}
\begin{abstract}
The fundamental relation betwen the slopes of diffraction cone in
single diffraction dissociation and elastic scattering has been
derived.
\end{abstract}



\vspace{0.5cm}
Not long ago we observed that the slope $b_{SD}$ of diffraction
cone in single diffraction dissociation $NN\rightarrow NX$ was
related to the effective interaction radius $R_0$ for the three-body
(three-nucleon) forces \cite{1}
\be
b_{SD}(s,M_X^2) = \frac{1}{2}R_0^2(\bar s,s'_0),\label{1}
\ee
\[
\bar s = 2(s + M_N^2) - M_X^2, \quad s'_0 = 2s_0,
\]
where $s_0$ is a scale defining unitarity saturation asymptotic in
hadron-hadron interaction. At the same time it was established that
the quantity $R_0^2$ was related to the structure of hadronic total
cross section in a physically clear and transparent form \cite{1}
(see also \cite{2,3,4})
\be
\sigma^{tot}(s) = 2\pi\left[B^{el}(s) +
R_0^2(s)\right]\left(1+\chi(s)\right), \label{2}
\ee
where $B^{el}$ is the slope of diffraction cone in elastic $NN$
scattering 
and 
\[
\chi(s) = O\left(\frac{1}{\sqrt{s}\ln^3s}\right),\quad s\rightarrow
\infty.
\]
This circumstance gives rise to the nontrivial consequences which
are discussed in this note.

Let us define the slope $B^{sd}$ of diffraction cone in a single
diffraction dissociation at the fixed point over the missing mass
\be
B^{sd}(s) = b_{SD}(s,M_X^2)\vert_{M_X^2 = 2M_N^2}.\label{3}
\ee
Now taking into account Eq. (\ref{2}), where the effective
interaction radius for three-body forces can be extracted from
\cite{5}
\be
R_0^2(2s,2s_0) = R_0^2(s,s_0) = \frac{1}{2\pi}\sigma^{tot}(s) -
B^{el}(s),\label{4}
\ee
and the equation
\be
\sigma^{el}(s) = \frac{\sigma^{tot}(s)^2}{16\pi B^{el}(s)},\quad
(\rho = 0),\label{5}
\ee
we come to the fundamental relation between the slopes in the single
diffraction dissociation and elastic scattering
\be
\fbox{$\displaystyle B^{sd}(s) = B^{el}(s)\left(4X -
\frac{1}{2}\right)$}\,,\label{6}
\ee
where
\be
X \equiv \frac{\sigma^{el}(s)}{\sigma^{tot}(s)}.\label{7}
\ee
The quantity $X$ has a clear physical meaning, it has been introduced
in the papers of C.N. Yang and his collaborators \cite{6,7}.

In paper \cite{8} we search for $X = 0.25$ at $\sqrt{s} = 1800\,
GeV$. Hence in that case we have  $B^{sd} = B^{el}/2$ which is
confirmed not so badly in the experimental measurements \cite{8}.

In the limit of black disk $(X = 1/2)$ we obtain
\be
B^{sd} = \frac{3}{2}B^{el},\label{8}
\ee
and
\be
B^{sd} = B^{el},\quad at\quad X = \frac{3}{8} = 0.375.\label{9}
\ee

So, we find that there is quite a nontrivial dynamics in the slopes
of diffraction cone in the single diffraction dissociation and
elastic scattering processes. In particular, we can study an
intriguing question on black disk limit not only in the measurements
of total hadronic cross sections compared with elastic ones but in
the measurements of the slopes in single diffraction dissociation
processes together with  elastic scattering ones. 

There is a more general formula which can be derived with account of
the real parts for the amplitudes. This formula  looks like
\be
\fbox{$\displaystyle B^{sd}(s) = B^{el}(s)\left(4X\frac{1 -
\rho_{el}(s)\rho_0(s)}{1 + \rho_{el}^2(s)} -
\frac{1}{2}\right)$}\,,\label{10}
\ee
where $\rho_0$ is defined in terms of three-body forces scattering
amplitude similar to $\rho_{el}$ \cite{9}. If $\rho_{el} = 0$ or
$\rho_0 = -\rho_{el}$ then we come to Eq. (\ref{6}). In the case
when $\rho_{el} \not= 0$, we can rewrite Eq. (\ref{10}) in the form
\be
\rho_0 = \frac{1}{\rho_{el}}\left[1 - \frac{1 +
\rho_{el}^2}{8X}\left(1 +
\frac{2B^{sd}}{B^{el}}\right)\right].\label{11}
\ee
Eq. (\ref{11}) can be used for the calculation of the new quantity
$\rho_0$. Anyway it seems that the measurements of real parts for the
amplitudes will play an important role in the future high energy
hadronic physics.

Equations (\ref{6},\ref{10}) can be rewritten in a unique
form
\be
\frac{Y}{Y'} + \frac{1}{2} = \alpha_{\varrho}X,\label{12}
\ee
where the quantity $Y$ has also been introduced in the above
mentioned papers of C.N. Yang and his collaborators \cite{6,7}
\be
Y = \frac{\sigma^{tot}}{16\pi B^{el}},\label{13}
\ee
$\alpha_{\varrho}$ is a known function of $\rho$'s (see Eq.
(\ref{10})) and we introduced a new dimensionless quantity $Y'$
\be
Y' = \frac{\sigma^{tot}}{16\pi B^{sd}}.\label{14}
\ee
It is obvious
\be
\frac{Y}{Y'} \equiv \frac{B^{sd}}{B^{el}}, \label{15}
\ee
and we have for the quantity $\alpha_{\varrho} = 4$ if $\rho_{el} =
0$ or $\rho_0 = -\rho_{el}$.

Equation (\ref{12}) represents the fundamental constraint on 
three dimensionless quantities $Y, Y'$ and $X$. It would be very
desirable to experimentally study this constraint.

%\newpage

\end{document}